\documentstyle[aps,prd,epsfig,preprint]{revtex}
\def\fun#1#2{\lower3.6pt
\vbox{\baselineskip0pt\lineskip.9pt
\ialign{$\mathsurround=0pt#1\hfill##\hfil$
\crcr#2\crcr\sim\crcr}}}
\begin{document}
\vspace{0.5in}
\title{\vskip-2.5truecm{\hfill \baselineskip 14pt 
{\hfill {{\small \hfill UT-STPD-2/00}}}\\
{{\small \hfill FISIST/2-2000/CFIF}}
\vskip .1truecm} 
\vspace{1.0cm}
\vskip 0.1truecm{\bf Yukawa Unification, $b\rightarrow s\gamma$
and Bino-Stau Coannihilation}}
\vspace{1cm}
\author{{M. E. G\'omez}$^{(1)}$\thanks{mgomez@gtae2.ist.utl.pt},
{G. Lazarides}$^{(2)}$\thanks{lazaride@eng.auth.gr} 
{and C. Pallis}$^{(2)}$\thanks{kpallis@gen.auth.gr}} 
\vspace{1.0cm}
\address{$^{(1)}${\it Centro de F\'{\i}sica das 
Interac\c{c}\~{o}es Fundamentais (CFIF),  
Departamento de F\'{\i}sica, \\ Instituto Superior T\'{e}cnico, 
Av. Rovisco Pais, 1049-001 Lisboa, Portugal.}} 
\address{$^{(2)}${\it Physics Division, School of Technology, 
Aristotle University of Thessaloniki,\\ 
Thessaloniki GR 540 06, Greece.}}
\maketitle

\vspace{.8cm}

\begin{abstract}
\baselineskip 12pt

\par
The minimal supersymmetric standard model with universal 
boundary conditions and ``asymptotic" Yukawa unification 
is considered. The full one-loop effective potential 
for radiative electroweak symmetry breaking as well as the 
one-loop corrections to the charged Higgs boson, $b$-quark 
and $\tau$-lepton masses are included. The CP-even Higgs 
boson masses are corrected to two-loops. The relic abundance 
of the lightest supersymmetric particle (bino) is calculated 
by including its coannihilations with the next-to-lightest 
supersymmetric particle (lightest stau) consistently with 
Yukawa unification. The branching ratio of 
$b\rightarrow s\gamma$ is evaluated by incorporating all 
the applicable next-to-leading order QCD corrections. The 
bino-stau coannihilations reduce the bino relic abundance 
below the upper bound from cold dark matter considerations 
in a sizable fraction of the parameter space allowed by 
$b\rightarrow s\gamma$ for $\mu>0$. Thus, the $\mu>0$ 
case, which also predicts an acceptable $b$-quark mass, 
is perfectly compatible with data.

\end{abstract}

\thispagestyle{empty}
\newpage
\pagestyle{plain}
\setcounter{page}{1}
\baselineskip 20pt

\par
It is well-known \cite{als} that the assumption that all three 
Yukawa couplings of the third family of quarks and leptons unify 
``asymptotically" (i.e., at the grand unified theory (GUT) mass 
scale $M_{GUT}\sim 10^{16}~{\rm{GeV}}$) naturally restricts 
the top quark mass to large values compatible with the present 
experimental data. Such a Yukawa unification can be obtained by 
embedding the minimal supersymmetric standard model (MSSM) in a 
supersymmetric (SUSY) GUT with a gauge group such as $SU(4)_c
\times SU(2)_L\times SU(2)_R$, $SO(10)$ or $E_{6}$ which 
contain $SU(4)_c$ and $SU(2)_R$. Assuming that the electroweak 
Higgs superfields $H_1$, $H_2$ and the third family right-handed 
quark superfields $t^c$, $b^c$ form $SU(2)_R$ doublets, we 
obtain \cite{pana} the ``asymptotic" Yukawa coupling relation 
$h_t=h_b$ and, hence, large $\tan\beta\approx m_{t}/m_{b}$.
Moreover, if the third generation quark and lepton $SU(2)_L$ 
doublets (singlets) $Q_3$ and $L_3$ ($b^c$ and $\tau^c$) 
form a $SU(4)_c$ 4-plet ($\bar 4$-plet) and the electroweak 
Higgs $H_1$ which couples to them is a $SU(4)_c$ singlet, we 
obtain $h_{\tau}=h_b$ and the successful ``asymptotic" mass 
relation $m_{\tau}=m_{b}$ follows.

\par
The simplest and most restrictive version of MSSM with gauge 
coupling unification is based on the assumption of radiative 
electroweak symmetry breaking with universal boundary conditions 
from gravity-mediated soft SUSY breaking. The tantalizing 
question is then whether this scheme is compatible with exact 
``asymptotic" unification of the three third family Yukawa 
couplings. A positive answer to this question would be very 
desirable since it would lead to a simple and highly predictive 
theory. This issue has been systematically studied in 
Ref.\cite{copw}. 

\par
A significant problem, which may be faced in trying to reconcile 
Yukawa unification and universal boundary conditions, is due to 
the generation of sizeable SUSY corrections to the $b$-quark 
mass \cite{copw,hall}. The sign of these corrections is 
opposite to the sign of the MSSM parameter $\mu$ (with the 
conventions of Ref.\cite{cdm}). As a consequence, for $\mu<0$, 
the tree-level value of $m_b$, which is 
predicted from Yukawa unification already near its experimental 
upper bound, receives large positive corrections which drive 
it well outside the allowed range. However, it  
should be noted that this problem arises in the simplest 
realization of this scheme. In complete models correctly
incorporating fermion masses and mixing, $m_b$ can receive
extra corrections which may make it compatible with experiment. 
Also, small GUT threshold corrections to gauge 
coupling unification can help to reduce $m_b$. (For a brief 
discussion of the possibilities to remedy the $m_b$ problem
encountered in the $\mu<0$ case and some relevant references 
see Ref.\cite{cdm}.) So, we do not consider this $b$-quark 
mass problem absolutely fatal for the $\mu<0$ case.

\par
Be that as it may, it is certainly interesting to examine the 
alternative scenario with $\mu>0$ too. The $b$-quark mass 
receives negative SUSY corrections and can easily be compatible 
with data in this case. This scheme, however, is severely
restricted by the recent experimental results \cite{cleo} on 
the inclusive decay $b\rightarrow s\gamma$ \cite{bsg}. It 
is well-known that the SUSY corrections to the inclusive 
branching ratio BR($b\rightarrow s\gamma$), in the case of 
the MSSM with universal boundary conditions, arise mainly 
from chargino loops and have the same 
sign with the parameter $\mu$. Consequently, these 
corrections interfere constructively with the contribution 
from the standard model (SM) including an extra electroweak 
Higgs doublet. However, this contribution is already bigger than 
the experimental upper bound on BR($b\rightarrow s\gamma$) 
for not too large values of the CP-odd Higgs boson mass $m_A$. 
As a result, in the present context with Yukawa unification and 
hence large $\tan\beta$, a lower bound on $m_A$ is obtained 
\cite{borzumati,bsg2} for $\mu>0$. On the contrary, for 
$\mu<0$, the SUSY corrections to BR($b\rightarrow s\gamma$)
interfere destructively with the SM plus extra Higgs doublet 
contribution yielding, in most cases, no restrictions on the 
parameters.

\par
An additional constraint results from the requirement that the 
relic abundance $\Omega_{LSP}~h^2$ of the lightest 
supersymmetric particle (LSP) in the universe does not exceed 
the upper limit on the cold dark matter (CDM) abundance implied 
by cosmological considerations ($\Omega_{LSP}$ is the present 
energy density of the LSPs over the critical energy density of 
the universe and $h$ is the present value of the Hubble constant 
in units of $100~\rm{km}~\rm{sec}^{-1}~\rm{Mpc}^{-1}$). Taking 
both the currently available cosmological models with 
zero/nonzero cosmological constant, which provide the best fits 
to all the data, as equally plausible alternatives for the 
composition of the energy density of the universe and accounting 
for the observational uncertainties, we obtain the restriction 
$\Omega_{LSP}~h^{2}\stackrel{_{<}}{_{\sim }}0.22$ (see 
Refs.\cite{cdm,lahanas}). Assuming that all the CDM in the 
universe is composed of LSPs, we further get 
$\Omega_{LSP}~h^{2}\stackrel{_{>}}{_{\sim }}0.09$.

\par
The LSP is normally the lightest neutralino ($\tilde\chi$). In 
the particular case of Yukawa unification, this neutralino turns 
out to be an almost pure bino. Its relic abundance 
has been estimated, for $\mu>0$, and shown \cite{borzumati,bsg2} 
to be well above unity, thereby overclosing the universe for all 
$m_A$'s permitted by $b\rightarrow s\gamma$. So, the 
combination of CDM considerations and the data on 
BR($b\rightarrow s\gamma$) seems to rule out the MSSM with 
$\mu>0$, Yukawa unification and radiative electroweak breaking 
with universal boundary conditions.

\par
It is important to note that, in Refs.\cite{borzumati,bsg2}, the
coannihilation \cite{coan} of the LSP with the next-to-lightest 
supersymmetric particle (NLSP) has been ignored and only the LSP
annihilation processes have been taken into account. It is 
well-known, however, that these coannihilations can be extremely 
important, if the mass of the NLSP is relatively close to the mass 
of the LSP, resulting to a considerable reduction of the LSP relic 
abundance \cite{cdm,drees,ellis}. The question then arises
whether, by employing LSP-NLSP coannihilation, one can succeed 
reducing $\Omega_{LSP}~h^{2}$ below 0.22 for some values of 
$m_A$ allowed by $b\rightarrow s\gamma$ in the $\mu>0$ case. 
This would revitalize a part of the available parameter space for 
$\mu>0$ saving the simple, elegant and predictive MSSM with 
universal boundary conditions and Yukawa unification even in its 
simplest realization (with no need of extra corrections to 
$m_b$). Although, in this parameter range, the sparticles would 
be quite massive (due to the lower bound on $m_A$ from 
$b\rightarrow s\gamma$) to be of immediate phenomenological 
interest, we would consider this as a very positive development.

\par
In this paper, we reconsider the constraints from $m_b$, 
$b\rightarrow s\gamma$ and the LSP relic density in the context 
of MSSM with universal boundary conditions and Yukawa unification 
by incorporating the LSP ($\tilde\chi$) and NLSP coannihilation 
in the calculation of $\Omega_{LSP}~h^{2}$. The NLSP turns out 
to be the lightest stau mass eigenstate $\tilde\tau_2$ and its 
coannihilation with bino has been studied in Ref.\cite{cdm} 
(or \cite{ellis}) for large (or small) $\tan\beta$. Although 
our analysis covers both signs of $\mu$, our main interest here 
is to see whether, for $\mu>0$, there exists a range of parameters 
where all these constraints are simultaneously satisfied.

\par
We will consider the MSSM with Yukawa unification described in 
detail in Ref.\cite{cdm} and closely follow the notation as 
well as the renormalization group (RG) and radiative electroweak 
symmetry breaking analysis of this reference. The only essential 
improvement here is the inclusion of the full one-loop radiative 
corrections to the effective potential for the electroweak 
symmetry breaking which have been evaluated in Ref.\cite{pierce} 
(Appendix E). We also incorporate the one-loop corrections to 
certain particle masses from the same reference and the two-loop 
corrections to the CP-even neutral Higgs boson masses (see below). 
In Ref.\cite{cdm}, we used a constant common SUSY threshold 
$M_S=1~{\rm{TeV}}$, where the RG-improved tree-level 
potential was minimized and the parameters $\mu$ and $m_A$ 
were evaluated. Also, the MSSM RG equations were 
replaced by the SM ones below the same scale $M_S$. Here, we 
use a variable common SUSY threshold $M_S=\sqrt{m_{\tilde t_1}
m_{\tilde t_2}}$, where $\tilde t_{1,2}$ are the two stop 
mass eigenstates. As it turns out, this does not make a very 
significant numerical difference since the values of $\mu$ and 
$m_A$ are found to be pretty stable to variations of $M_S$. 
Despite this fact, it is more appropriate to use a variable 
SUSY threshold here because of the wide variation of the 
sparticle spectrum encountered. This variation appears since, 
in this work, we consider not only relatively small but also 
quite large values of $m_A$. Finally, note that our choice 
of the SUSY threshold $M_S$ minimizes the size of the one-loop 
corrections to $\mu$ and $m_A$ and, thus, the errors in the 
determination of these parameters.   

\par
We take universal soft SUSY breaking terms at $M_{GUT}$, i.e., a 
common mass for all scalar fields $m_0$, a common gaugino mass 
$M_{1/2}$ and a common trilinear scalar coupling $A_0$, which 
we put equal to zero (we will discuss later the influence of 
non-zero $A_0$'s). Our effective theory below $M_{GUT}$ then 
depends on the parameters ($\mu_0=\mu(M_{GUT})$)
\[
m_0,\ M_{1/2},\ \mu_0,\ \alpha_G,\ M_{GUT},\ h_{0},\ 
\tan\beta~,  
\]
where $\alpha_G=g_G^2/4\pi$ ($g_G$ being the GUT gauge coupling 
constant) and $h_0$ is the common top, bottom and tau Yukawa 
coupling constant at $M_{GUT}$. The values of $\alpha_G$ and 
$M_{GUT}$ are obtained as described in Ref.\cite{cdm}.

\par
\begin{figure}
\epsfig{figure=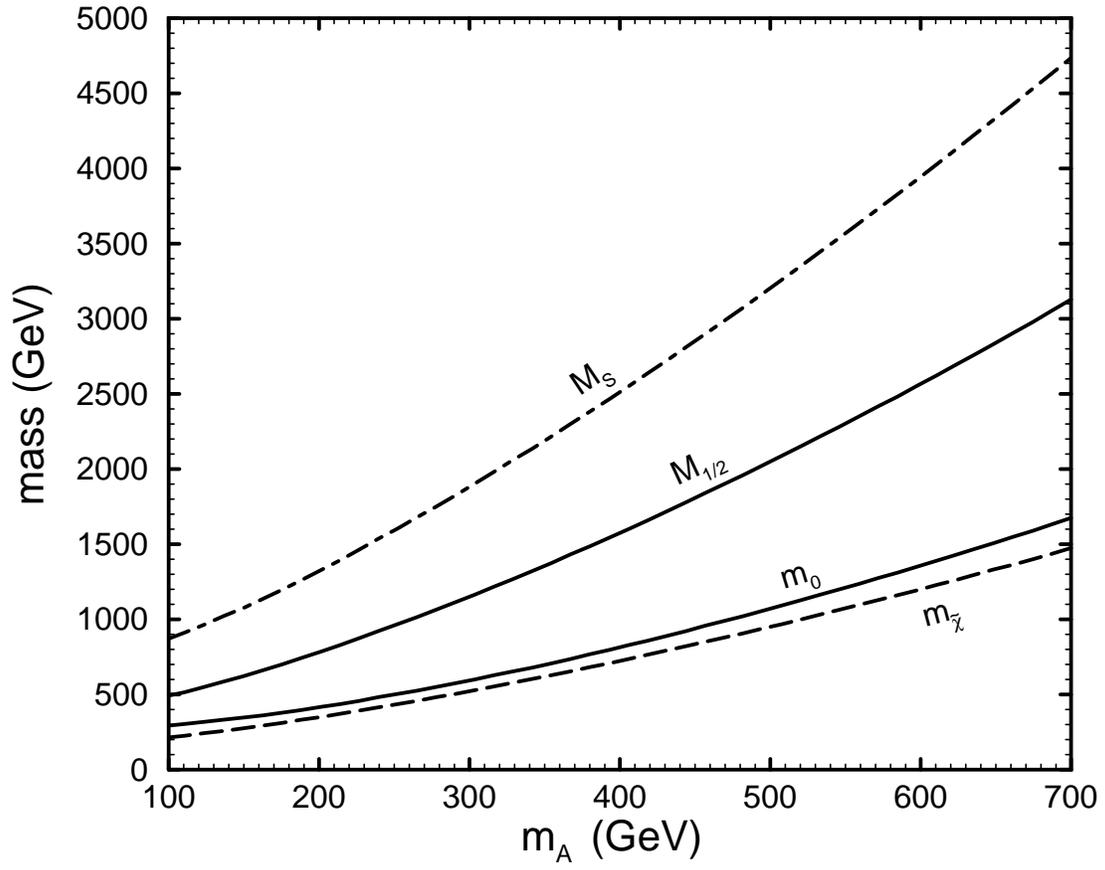,height=4.64in,angle=0}
\medskip
\caption{The values of $m_{\tilde\chi}$, $m_0$, $M_{1/2}$ 
and $M_S$ as functions of $m_A$ for $\mu>0$, $A_0=0$ and 
$m_{\tilde\tau_2}=m_{\tilde\chi}$. These values are affected 
very little by changing the sign of $\mu$.
\label{ms}}
\end{figure}

It was pointed out \cite{copw} that, for every $m_A$, the 
requirement of successful radiative electroweak symmetry 
breaking with Yukawa unification implies a relation among the GUT 
parameters $m_0$ and $M_{1/2}$. This relation combined   
with any given ratio of the masses of the two lightest SUSY 
particles (lightest neutralino and stau) leads \cite{cdm} to a 
complete determination of the values of $m_0$ and $M_{1/2}$ and 
hence the whole SUSY spectrum for every value of $m_A$ (see also 
Ref.\cite{anant}). We will thus use $m_A$ and 
$\Delta_{\tilde\tau_2}=(m_{\tilde\tau_2}-m_{\tilde\chi})/
m_{\tilde\chi}$ as our basic independent parameters. For reasons 
to become obvious later, we concentrate here on the limiting case 
with $m_{\tilde\tau_2}=m_{\tilde\chi}$ 
($\Delta_{\tilde\tau_2}=0$) where the LSP and NLSP 
coannihilation is \cite{cdm} most efficient. The values of $m_0$ 
and $M_{1/2}$ can then be found as functions of $m_A$ and are 
depicted in Fig.\ref{ms}, for $\mu>0$, together with the LSP mass 
($m_{\tilde\chi}$) and the SUSY threshold mass parameter $M_S$.
Note that these mass parameters are affected very little by changing 
the sign of $\mu$.

\par
We find that the effect of including the full one-loop effective 
scalar potential from Ref.\cite{pierce} is not very significant. 
In particular, the correction to the tree-level value of $m_A$ is 
found to range from about $-7.5\%$ to about $1.5\%$ for $m_A$ 
between 100 and $700~{\rm{GeV}}$ with only a small dependence 
on the sign of $\mu$. Also, the one-loop radiative correction to 
the tree-level value of $\mu$ turns out to be of the same sign as 
$\mu$ and less than about $3.5\%$ in the same range of $m_A$. 

\par
The values of the unified Yukawa coupling constant $h_{0}$ at 
$M_{GUT}$ and of $\tan\beta$ at $M_S$ are estimated by 
using the running top quark mass at $m_t$, 
$m_t(m_t)=166~{\rm{GeV}}$, and the running tau lepton mass at 
$m_Z$, $m_\tau(m_Z)=1.746~{\rm{GeV}}$. We also incorporate 
the SUSY threshold correction to $m_\tau(M_S)$ from the 
approximate formula of Ref.\cite{pierce}. This correction arises 
mainly from chargino/tau sneutrino 
loops, is almost $m_A$-independent and has the same sign as 
$\mu$. It is about $8\%$, for $\mu>0$, leading to a value of 
$\tan\beta=55.4-54.5$ for $m_A=100-700~{\rm{GeV}}$, while, 
for $\mu<0$, we find a correction of about $-7\%$ and 
$\tan\beta=47.8-46.9$ in the same range of $m_A$.

\par
The tree-level values which we find for $m_b(m_Z)$ are quite close 
to its experimental upper bound \cite{mb}:
\[
m_b(m_Z)=2.67 \pm 0.50~{\rm GeV}.
\]
The SUSY correction \cite{copw,hall} to the bottom quark mass is 
known to be very large for models with Yukawa unification. This 
correction originates mainly from squark/gluino and squark/chargino 
loops and has sign opposite to the one of $\mu$ (in our convention). 
Thus, for $\mu<0$, the corrected $m_b(m_Z)$ will certainly be 
outside the experimentally allowed range. For $\mu>0$, however, 
this large negative correction may easily make $m_b(m_Z)$ 
compatible with data. Indeed, using the approximate formula of
Ref.\cite{pierce}, we find that, for $\mu<0$, the correction 
is about $27.2\%-23.8\%$ for $m_A=100-700~{\rm{GeV}}$, which 
added to a tree-level value of $m_b(m_Z)\approx 3.41~{\rm{GeV}}$ 
leads to an unacceptably large $m_b(m_Z)$. For $\mu>0$, the 
tree-level value of $m_b(m_Z)$ is equal to about 
$3.13~{\rm{GeV}}$ and the SUSY correction, to be subtracted from 
it, about $26.6\%-24.7\%$ for $m_A=100-700~{\rm{GeV}}$. 
The resulting bottom quark mass is then perfectly acceptable in 
this case. Note that the variation of the tree-level value of 
$m_b(m_Z)$ with the sign of $\mu$ is due to the SUSY corrections 
to $m_\tau$ considered in calculating $\tan\beta$.   

\par
We incorporate in our calculation the two-loop corrections to 
the CP-even neutral Higgs boson mass matrix by employing the 
program {\it FeynHiggsFast} \cite{fh}. The tree-level mass 
of the lightest neutral CP-even Higgs boson is very close to 
$m_Z$ for all $m_A$'s considered here. Two-loop corrections, 
however, increase its value, $m_h$, considerably. The lower 
experimental bound on $m_h$, say equal to $105~{\rm{GeV}}$, 
corresponds to $m_A\approx 104~{\rm{GeV}}$. The value of 
$m_h$ increases rapidly with $m_A$ reaching about 
$115~{\rm{GeV}}$ at $m_A=120~{\rm{GeV}}$. After this, 
the growth of $m_h$ slows down drastically and this mass 
soon enters into a plateau with $m_h\approx 122~{\rm{GeV}}$. 
The difference between the two-loop and the tree-level value 
of the heavier CP-even neutral Higgs boson mass $m_H$ is 
insignificant for values of $m_A$ bigger than about 
$120~{\rm{GeV}}$. For smaller $m_A$'s, this difference 
increases reaching about $10\%$ of the tree-level value of 
$m_H$ at $m_A\approx 100~{\rm{GeV}}$. Finally, we also 
include the one-loop corrections to the charged Higgs boson 
mass using the formalism described in the appendices of 
Ref.\cite{pierce} but without the neutralino and chargino 
contributions since, as it turns out, these contributions are 
relatively unstable as we change the scale $M_S$. Presumably, 
higher order corrections will alleviate this instability. It 
is, certainly, reassuring for our procedure that similar 
results are obtained by using the approximate formula of 
Ref.\cite{carena}. We find that one-loop corrections increase 
the tree-level value of the charged Higgs boson mass 
$m_{H^+}$. This increase ranges from $7\%$ to $5\%$ of the 
tree-level value as $m_A$ changes from 100 to 
$700~{\rm{GeV}}$.

\par
To study the constraints imposed by $b\rightarrow s\gamma$ 
on the parameter space of our model, we follow the analysis of 
Ref.\cite{kagan}. We consider the SM contribution to the 
inclusive branching ratio BR($b\rightarrow s\gamma$) from 
a loop with a $W$-boson and top quark ($t$), the contribution 
from loops with charged Higgs bosons (with mass corrected to 
one-loop) and $t$ and the dominant SUSY contribution arising 
from loops with charginos and stop quarks. The SM contribution,
which is factorized out in the formalism of Ref.\cite{kagan},
includes the next-to-leading order (NLO) QCD \cite{nlosm} 
and the leading order (LO) QED \cite{kagan,loqed} corrections.
The NLO QCD corrections \cite{nlohiggs} to the charged Higgs 
boson contribution are taken from the first paper in 
Ref.\cite{nlohiggs}. The SUSY contribution is evaluated by 
including only the LO QCQ corrections \cite{bsg,nlosusy} 
using the formulae in Ref.\cite{nlosusy}. NLO QCD corrections 
to the SUSY contribution have also been discussed in 
Ref.\cite{nlosusy}, but only under certain very restrictive 
conditions which never hold in our case since the  
lightest stop quark mass is comparable to the masses of the 
other squarks and the gluinos. Also, the charginos are pretty 
heavy. We, thus, do not include these corrections in our 
calculation. (Moreover, as pointed out in Ref.\cite{bmu}, 
the available NLO QCD corrections to the SUSY contribution 
are not applicable to models with large values of 
$\tan\beta$, which is our case here.) The results, evaluated 
with central values of the input parameters and the 
renormalization and matching scales, are depicted in 
Fig.\ref{bsg} for both signs of $\mu$, $A_0=0$ and 
$m_{\tilde\tau_2}=m_{\tilde\chi}$ (see below). The 
charged Higgs contribution, added to the one of the SM, raises 
the predicted value of BR($b\rightarrow s\gamma$) above the 
experimental upper bound $4.5\times 10^{-4}$ \cite{cleo} 
for not too large values of $m_A$. The SUSY contribution, 
which becomes less important as $m_A$ increases, interferes 
constructively or destructively with the other two 
contributions for $\mu$ positive or negative respectively.

\par
We see from Fig.\ref{bsg} that the SM plus charged Higgs 
contribution decreases as $m_A$ (or $m_{H^+}$) increases 
and enters into the experimentally allowed range at 
$m_{A}\approx 295~{\rm{GeV}}$ corresponding to 
$m_{H^+}\approx 318~{\rm{GeV}}$ and 
$M_S\approx 1851~{\rm{GeV}}$. For $\mu<0$, inclusion of 
the SUSY contribution makes the BR($b\rightarrow s\gamma$) 
compatible with data for all values of $m_A$ explored here and 
no useful restrictions on the parameter space are obtained for 
$m_{\tilde\tau_2}=m_{\tilde\chi}$ and $A_0=0$. (We may, 
though, obtain restrictions for $m_{\tilde\tau_2}$'s higher 
than $m_{\tilde\chi}$ and/or $A_0\neq 0$.) For $\mu>0$, 
however, the SUSY contribution increases the discrepancy 
between the predicted value of this branching ratio and the 
data. The upper experimental limit on 
BR($b\rightarrow s\gamma$) ($\approx 4.5\times 10^{-4}$) 
is reached at $m_{A}\approx 385~{\rm{GeV}}$ corresponding to 
$m_{\tilde\chi}\approx 694~{\rm{GeV}}$, $m_0\approx 
781~{\rm{GeV}}$, $M_{1/2}\approx 1512~{\rm{GeV}}$ and 
$M_S\approx 2418~{\rm{GeV}}$. We conclude that, for $\mu>0$, 
$A_0=0$ and $m_{\tilde\tau_2}=m_{\tilde\chi}$, $m_{A}$ 
should be greater than about $385~{\rm{GeV}}$ for satisfying 
the constraints from the $b\rightarrow s\gamma$ process. It 
is important to observe that, for values of $m_{\tilde\tau_2}$ 
higher than $m_{\tilde\chi}$, charginos become even heavier 
and the lower bound on $m_A$ decreases slightly but it can never 
become smaller than the bound ($295~{\rm{GeV}}$) from the SM 
plus charged Higgs contribution.

\par
The lower bound on $m_A$ found by using central values of 
the input parameters can be considerably reduced if the 
theoretical uncertainties entering into the calculation 
of BR($b\rightarrow s\gamma$) are taken into account. 
These uncertainties originating from the experimental 
errors in the input parameters and the ambiguities in the 
renormalization and matching scales are known to be quite 
significant. The SM contribution alone, which is factorized 
out, generates an uncertainty of about $\pm 10\%$. The 
charged Higgs and SUSY contributions can only increase this 
uncertainty. In particular, the SUSY prediction for $\mu>0$ 
should not be a line in Fig.\ref{bsg} but rather a band 
with the appropriate error margin. Consequently, it should 
intersect the upper experimental bound line at a segment 
rather than at a point. This segment is found to be from 
about 300 to $510~{\rm{GeV}}$ if only the ambiguities 
from the SM contribution are taken into account. We see 
that the lower bound on $m_A$ is reduced from about 385 
to about $300~{\rm{GeV}}$ at least. Inclusion of the 
errors from the charged Higgs and chargino contributions 
can reduce this bound even further. However, we believe 
that they cannot be reliably calculated at the moment 
since the NLO QCD corrections to the SUSY contribution 
are not known in our case. In any case, the lower bound 
of $300~{\rm{GeV}}$ on $m_A$ is more than adequate 
for our purpose here, which is to revitalize the MSSM 
with Yukawa unification, universal boundary conditions 
and $\mu>0$. Note that, for values of 
$m_{\tilde\tau_2}$ higher than $m_{\tilde\chi}$ 
this lower bound on $m_A$ decreases slightly but it 
can never become smaller than the corresponding bound 
($\approx 200~{\rm{GeV}}$) derived from the SM plus 
charged Higgs contribution.

\par
\begin{figure}
\epsfig{figure=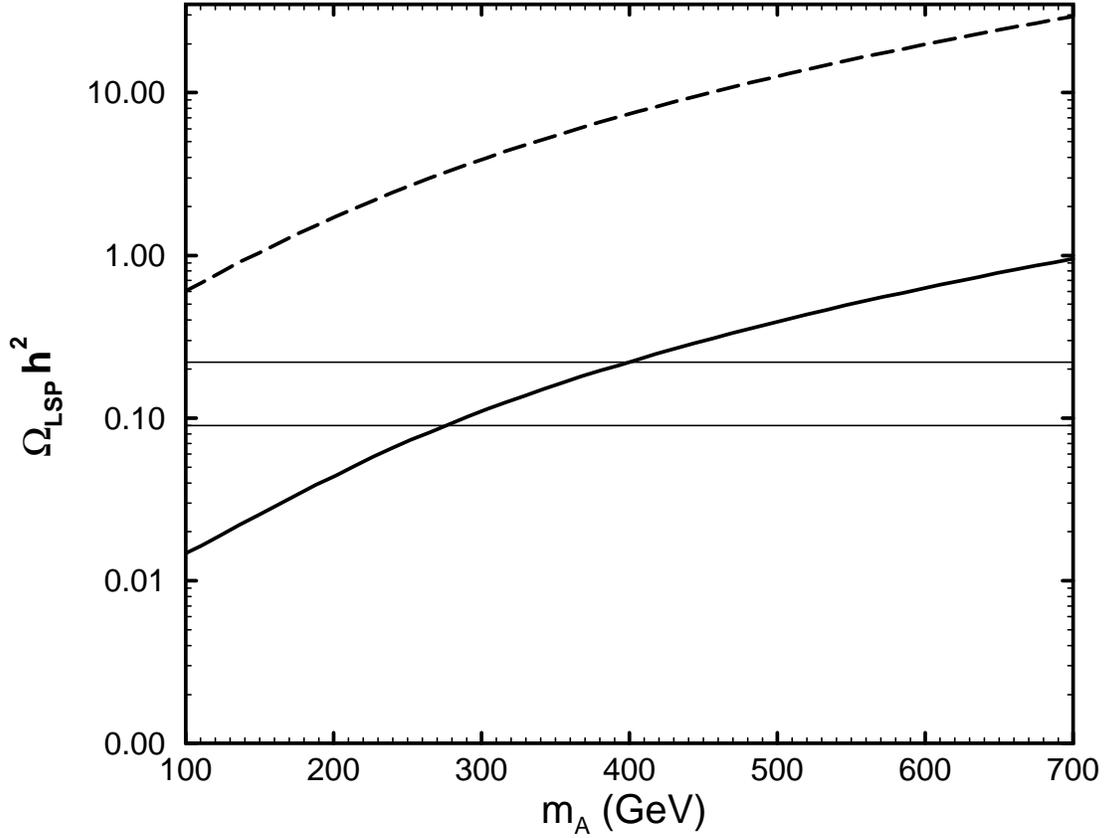,height=4.47in,angle=0}
\medskip
\caption{The LSP relic abundance $\Omega_{LSP}~h^2$ as function of 
$m_A$ in the limiting case $m_{\tilde\tau_2}=m_{\tilde\chi}$ and 
for $\mu>0$, $A_0=0$. The solid line includes coannihilation of 
$\tilde\tau_2$ and $\tilde\chi$, while the dashed line is obtained 
by only considering the LSP annihilation processes. These results are 
affected very little by changing the sign of $\mu$. The limiting lines 
at $\Omega_{LSP}~h^2=0.09$ and 0.22 are also included.
\label{lspp}}
\end{figure}

The relic abundance of the LSP ($\tilde\chi$), which is a 
nearly pure bino in our model, can be calculated by employing 
the analysis of Ref.\cite{cdm} which is appropriate for 
Yukawa unification and, thus, large $\tan\beta$. The inclusion 
of coannihilation effects of the LSP with the NLSP, which is the 
lightest stau mass eigenstate ($\tilde\tau_2$), is of crucial 
importance. These effects can reduce considerably the LSP relic 
abundance $\Omega_{LSP}~h^2$ so as to have a chance to 
satisfy the upper bound 0.22, derived from CDM considerations, 
for values of $m_A$ consistent with the constraints from 
$b\rightarrow s\gamma$. In order to achieve maximal 
coannihilation and thus obtain the strongest possible reduction 
of $\Omega_{LSP}~h^2$, we consider the limiting case 
$m_{\tilde\tau_2}=m_{\tilde\chi}$ (see Ref.\cite{cdm}). 
The relevant coannihilation processes and Feynman graphs together 
with the corresponding analytical expressions can be found in 
Ref.\cite{cdm}. Our computation here is, however, more accurate 
since it includes the one-loop corrections to the parameter 
$\mu$, the $\tau$-lepton mass, $m_A$, $m_{H^+}$ and the 
two-loop corrections to the CP-even neutral Higgs boson masses. 
The results are shown in Fig.\ref{lspp} for $\mu>0$ and 
$A_0=0$. For comparison, we also include the value of 
$\Omega_{LSP}~h^2$ obtained by ignoring coannihilation effects. 
Note that these results remain essentially unaltered by changing 
the sign of $\mu$.

\par
From Fig.\ref{lspp}, one readily finds that the CDM constraint
on the LSP relic density $0.09\stackrel{_{<}}{_{\sim }}
\Omega_{LSP}~h^{2}\stackrel{_{<}}{_{\sim }}0.22$ is 
satisfied for $m_A$'s between about $275$ and 
$400~{\rm{GeV}}$ for $\mu$ positive, $A_0=0$ and 
$m_{\tilde\tau_2}=m_{\tilde\chi}$. Of course, the upper 
bound on $m_A$ is more general than the lower one since 
it hold even if the CDM does not solely consist of LSPs. 
Note that the upper bound on $m_A$ 
($\approx 400~{\rm{GeV}}$) in the $\mu>0$ case corresponds 
to $m_{\tilde\chi}\approx 724~{\rm{GeV}}$, 
$m_0\approx 815~{\rm{GeV}}$, 
$M_{1/2}\approx 1575~{\rm{GeV}}$ and 
$M_S\approx 2513~{\rm{GeV}}$.

\par
It is obvious, from Fig.\ref{lspp}, that the reduction of 
$\Omega_{LSP}~h^2$ caused by the coannihilation effects 
is dramatic and can bring the LSP relic abundance below 
0.22 for $m_A$'s in the allowed range from 
$b\rightarrow s\gamma$ considerations in the $\mu>0$ 
case with $A_0=0$ and $m_{\tilde\tau_2}=m_{\tilde\chi}$. 
Indeed, at $m_A\approx 300~{\rm{GeV}}$, which is its 
lower bound if a $10\%$ theoretical error is allowed in 
BR($b\rightarrow s\gamma$), 
$\Omega_{LSP}~h^2\approx 0.112$ (or 3.92 with no 
coannihilation) and increases with $m_A$ reaching 0.22 
(or 7.4 with no coannihilation) at 
$m_A\approx 400~{\rm{GeV}}$, where ($90\%$ of) the 
central value of BR($b\rightarrow s\gamma$) is about 
($4\times 10^{-4}$) $4.44\times 10^{-4}$. We see
that, for $\Delta_{\tilde\tau_2}=0$, $A_0=0$ and 
$\mu>0$, there exists a range of $m_A$ (at least between 
300 and $400~{\rm{GeV}}$) where both the constraints 
from $b\rightarrow s\gamma$ and CDM can be satisfied. 
Note that, even with the central value of 
BR($b\rightarrow s\gamma$), this range does not 
disappear. It only shrinks to the interval between 385 and 
$400~{\rm{GeV}}$. The value of $\Omega_{LSP}~h^2$
with (without) coannihilation at 
$m_A\approx 385~{\rm{GeV}}$ is about 0.205 (6.85).

\par
So far we concentrated in the limiting case 
$\Delta_{\tilde\tau_2}=0$ where the coannihilation 
effects are more efficient and we took, for simplicity, 
$A_0=0$. We will now briefly discuss the effect of allowing 
general values of these quantities. Obviously, for any given 
$m_A$, the sparticles become heavier as we increase 
$\Delta_{\tilde\tau_2}$. The main effect of this is that 
coannihilation quickly faints away and the upper bound on 
$m_A$ from CDM considerations rapidly decreases. As a 
consequence, there exists an upper bound on the parameter 
$\Delta_{\tilde\tau_2}$ beyond which the allowed range 
of $m_A$ disappears. Positive values of $A_0$ lead to 
heavier sparticle masses and, thus, to an increase of 
$\Omega_{LSP}~h^2$ and a slight decrease of
BR($b\rightarrow s\gamma$). The allowed range of $m_A$ 
again disappears above a positive value of $A_0$. Negative 
$A_0$'s bigger than about $-0.5M_{1/2}$ (generally) 
produce an insignificant decrease in the sparticle masses 
and $\Omega_{LSP}~h^2$. Lower negative values of $A_0$, 
however, lead again to an increase of the sparticle masses
and $\Omega_{LSP}~h^2$, which means that a negative 
lower bound on $A_0$ must also exist.

\par
We will not undertake here the difficult task of constructing 
the region in the $m_A$, $\Delta_{\tilde\tau_2}$, $A_0$ 
space which is consistent with the constraints from 
$b\rightarrow s\gamma$ and CDM considerations. This 
would require not only a detailed study of the theoretical 
uncertainties in the calculation of 
BR($b\rightarrow s\gamma$), but also inclusion of the
uncertainties associated with the particular implementation 
of the radiative electroweak symmetry breaking, the 
RG analysis and the radiative corrections to various particle 
masses. These uncertainties propagated to the sparticle 
spectrum, $\Omega_{LSP}~h^2$ and 
BR($b\rightarrow s\gamma$) can only widen the allowed 
region in the parameter space. Our main conclusion, which is 
the viability of MSSM with Yukawa unification, universal 
boundary conditions and $\mu>0$, can only be further 
strengthened by including these errors.

\par
In summary, we have considered the MSSM based on radiative 
electroweak symmetry breaking with universal boundary 
conditions and assumed unification of all three third family 
Yukawa couplings at the GUT scale. We employed the full 
one-loop effective potential for electroweak symmetry breaking 
as well as the one-loop corrections to the charged Higgs boson, 
$b$-quark and $\tau$-lepton masses. Also, two-loop 
corrections to the CP-even Higgs boson masses were taken. We 
imposed the constraints from $b\rightarrow s\gamma$ and 
CDM in the universe by carefully including in the LSP relic 
abundance calculation the LSP (bino) and NLSP (lightest stau) 
coannihilation effects in the case of Yukawa unification 
(and, thus, large $\tan\beta$). The calculation of the 
branching ratio of $b\rightarrow s\gamma$ incorporates all 
the applicable NLO QCD and LO QED corrections and some of 
its theoretical errors were taken into account. We found that 
bino-stau coannihilation drastically reduces the LSP relic 
density and succeeds to bring it below the CDM upper bound 
for $m_A$'s which are allowed by $b\rightarrow s\gamma$ in 
the $\mu>0$ case. This, combined with the fact that, for 
$\mu>0$, the bottom quark mass after SUSY corrections is 
experimentally acceptable, shows that the simple, elegant and 
restrictive version of MSSM with Yukawa unification and 
universal boundary conditions can be perfectly viable. It is 
important to note that, without bino-stau coannihilation, this 
model was excluded.  

\vspace{0.5cm}
We would like to thank M. Drees and R. Zhang for useful discussions. 
This work was supported by the European Union under TMR contract 
No. ERBFMRX--CT96--0090 and the Greek Government research grant 
PENED/95 K.A.1795. One of us (C. P.) thanks the Greek State 
Scholarship Institution (I. K. Y) for support.

\def\ijmp#1#2#3{{ Int. Jour. Mod. Phys. }{\bf #1~}(#2)~#3}
\def\pl#1#2#3{{ Phys. Lett. }{\bf B#1~}(#2)~#3}
\def\zp#1#2#3{{ Z. Phys. }{\bf C#1~}(#2)~#3}
\def\prl#1#2#3{{ Phys. Rev. Lett. }{\bf #1~}(#2)~#3}
\def\rmp#1#2#3{{ Rev. Mod. Phys. }{\bf #1~}(#2)~#3}
\def\prep#1#2#3{{ Phys. Rep. }{\bf #1~}(#2)~#3}
\def\pr#1#2#3{{ Phys. Rev. }{\bf D#1~}(#2)~#3}
\def\np#1#2#3{{ Nucl. Phys. }{\bf B#1~}(#2)~#3}
\def\npps#1#2#3{{ Nucl. Phys. (Proc. Sup.) }{\bf B#1~}(#2)~#3}
\def\mpl#1#2#3{{ Mod. Phys. Lett. }{\bf #1~}(#2)~#3}
\def\arnps#1#2#3{{ Annu. Rev. Nucl. Part. Sci. }{\bf
#1~}(#2)~#3}
\def\sjnp#1#2#3{{ Sov. J. Nucl. Phys. }{\bf #1~}(#2)~#3}
\def\jetp#1#2#3{{ JETP Lett. }{\bf #1~}(#2)~#3}
\def\app#1#2#3{{ Acta Phys. Polon. }{\bf #1~}(#2)~#3}
\def\rnc#1#2#3{{ Riv. Nuovo Cim. }{\bf #1~}(#2)~#3}
\def\ap#1#2#3{{ Ann. Phys. }{\bf #1~}(#2)~#3}
\def\ptp#1#2#3{{ Prog. Theor. Phys. }{\bf #1~}(#2)~#3}
\def\plb#1#2#3{{ Phys. Lett. }{\bf#1B~}(#2)~#3}
\def\apjl#1#2#3{{ Astrophys. J. Lett. }{\bf #1~}(#2)~#3}
\def\n#1#2#3{{ Nature }{\bf #1~}(#2)~#3}
\def\apj#1#2#3{{ Astrophys. Journal }{\bf #1~}(#2)~#3}
\def\anj#1#2#3{{ Astron. J. }{\bf #1~}(#2)~#3}
\def\mnras#1#2#3{{ MNRAS }{\bf #1~}(#2)~#3}
\def\grg#1#2#3{{ Gen. Rel. Grav. }{\bf #1~}(#2)~#3}
\def\s#1#2#3{{ Science }{\bf #1~}(19#2)~#3}
\def\baas#1#2#3{{ Bull. Am. Astron. Soc. }{\bf #1~}(#2)~#3}
\def\ibid#1#2#3{{ ibid. }{\bf #1~}(19#2)~#3}
\def\cpc#1#2#3{{ Comput. Phys. Commun. }{\bf #1~}(#2)~#3}
\def\astp#1#2#3{{ Astropart. Phys. }{\bf #1~}(#2)~#3}
\def\epj#1#2#3{{ Eur. Phys. J. }{\bf C#1~}(#2)~#3}

\newpage

\pagestyle{empty}

\begin{figure}
\epsfig{figure=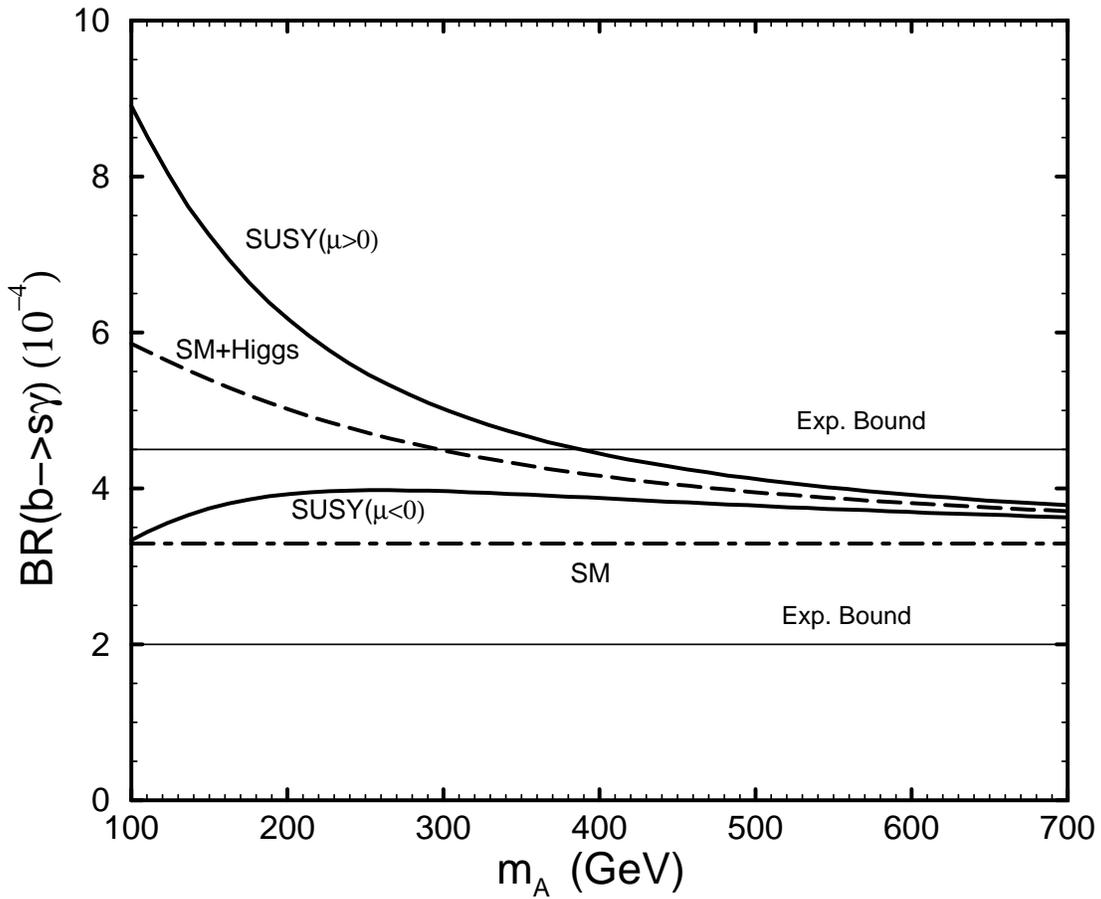,height=4.79in,angle=0}
\medskip
\caption{The central value of the SUSY inclusive 
BR($b\rightarrow s\gamma$) as function of $m_A$ for both signs 
of $\mu$, $A_0=0$ and $m_{\tilde\tau_2}=m_{\tilde\chi}$. 
The contributions from the SM and the SM plus charged Higgs boson 
(SM+Higgs) as well as the experimental bounds on 
BR($b\rightarrow s\gamma$), $2\times 10^{-4}$ and 
$4.5\times 10^{-4}$, are also indicated.    
\label{bsg}}
\end{figure}

\end{document}